\documentclass[11pt]{article} % For LaTeX2e

\usepackage[margin=1in]{geometry}

\usepackage{times}
\usepackage{pdfsync}
\usepackage{latexsym, epic, eepic,graphics,graphicx}
\usepackage{amsmath, amssymb, amsthm}
\usepackage[square,comma,numbers,sort&compress]{natbib}
\usepackage{color}
\usepackage{url}

\title{An Optimization-Based Framework for Automated Market-Making}
\author{
Jacob Abernethy\\
EECS Department\\
University of California, Berkeley\\
\texttt{\small jake@cs.berkeley.edu}
\and 
Yiling Chen\\
School of Engineering and Applied Sciences\\
Harvard University\\
\texttt{\small yiling@eecs.harvard.edu}
\and
\ \\
Jennifer Wortman Vaughan\\
Computer Science Department\\
University of California, Los Angeles\\
\texttt{\small jenn@cs.ucla.edu}}
\date{}

\newcommand{\ignore}[1]{}
\definecolor{darkgreen}{rgb}{0,0.5,0}
\definecolor{darkred}{rgb}{0.7,0,0}
\newcommand{\jenn}[1]{\textcolor{darkred}{[JWV: #1]}}
\newcommand{\yc}[1]{\textcolor{blue}{[YC: #1]}} 

\newcommand{\argmax}{\mathop{\rm argmax}}

\newcommand{\reals}{\mathbb{R}}

\newtheorem{theorem}{Theorem}
\newtheorem{proposition}{Proposition}
\newtheorem{definition}{Definition}
\newtheorem{lemma}{Lemma}

\newtheorem{condition}{Condition}
\renewcommand{\qed}{\hfill $\framebox(6,6){}$}

\newcommand{\e}{{\textrm{e}}}

\newcommand{\simp}{{\Delta}}

\newcommand{\squishlist}{
   \begin{list}{$\bullet$}
    { \setlength{\itemsep}{0pt}      \setlength{\parsep}{4pt}
      \setlength{\topsep}{4pt}       \setlength{\partopsep}{0pt}
     \setlength{\leftmargin}{2em} \setlength{\labelwidth}{1.5em}
      \setlength{\labelsep}{0.5em} } }
\newcommand{\squishend}{  \end{list}  }

\def\eb{\mathbf{e}}
\def\x{\mathbf{x}}
\def\u{\mathbf{u}}
\def\p{\mathbf{p}}
\def\q{\mathbf{q}}
\def\r{\mathbf{r}}
\def\1{\mathbf{1}}
\def\0{\mathbf{0}}
\def\m{\mathbf{m}}

\def\z{\mathbf{z}}
\def\e{\mathrm{e}}

\def\E{\mathbb{E}}

\def\reals{\mathbb{R}}

\def\M{{\mathrm{M}}}
\def\H{\mathcal{H}}
\def\HM{\mathcal{H}(\M)}

\def\dom{\text{dom}}
\def\y{\mathbf{y}}

\def\O{\mathcal{O}}
\def\o{\mathfrak{o}}
\def\cost{\mathtt{Cost}}
\def\rhob{\boldsymbol{\rho}}

\def\relint{\textnormal{relint}}
\def\aff{\textnormal{aff}}
\begin{document}

\maketitle

\begin{abstract} 
  Building on ideas from online convex optimization, we propose a general framework for the design of efficient securities markets over very large outcome spaces. The challenge here is computational. In a complete market, in which one security is offered for each outcome, the market institution can not efficiently keep track of the transaction history or calculate security prices when the outcome space is large. The natural solution is to restrict the space of securities to be much smaller than the outcome space in such a way that securities can be priced efficiently. Recent research has focused on searching for spaces of securities that can be priced efficiently by existing market mechanisms designed for operating complete markets. While there have been some successes, much of this research has led to hardness results.

In this paper, we take a drastically different approach. We start with an arbitrary space of securities with bounded payoff, and establish a framework to design markets tailored to this space. We prove that any market satisfying a set of intuitive conditions must price securities via a convex potential function and that the space of reachable prices must be precisely the convex hull of the security payoffs. We then show how the convex potential function can be defined in terms of an optimization over the convex hull of the security payoffs. The optimal solution to the optimization problem gives the security prices. Using this framework, we provide an efficient market for predicting the landing location of an object on a sphere. In addition, we show that we can relax our ``no-arbitrage'' condition to design a new efficient market maker for 
pair betting, which is known to be \#P-hard to price using existing mechanisms. This relaxation also allows the market maker to charge transaction fees so that the depth of the market can be dynamically increased as the number 
of trades increases.  
 \end{abstract}

\newpage

\section{Introduction} 

{\em Securities markets} play a fundamental role in economics and finance.
% for resource allocation under uncertainty. 
A securities market offers a set of {\em contingent securities} whose payoffs each depend on the future state of the world. For example, an Arrow-Debreu security pays \$1 if a particular state of the world is reached and \$0 otherwise~\cite{Arrow:64,Arrow:70}. Consider an Arrow-Debreu security that will pay off in the event that a category 4 or higher hurricane passes through Florida in 2011. A Florida resident who is worried about his home being damaged might buy this security as a form of insurance to hedge his risk; if there is a hurricane powerful enough to damage his home, he will be compensated. Additionally, a risk neutral trader who has reason to believe that the probability of a category 4 or higher hurricane landing in Florida in 2011 is $p$ should be willing to buy this security at any price below $p$ (or sell it at any price above $p$) to capitalize his information. For this reason, the market price of the security can be viewed as the traders' collective estimate of how likely it is that a powerful hurricane will occur. Securities markets thus have dual functions: risk allocation and information aggregation. Insurance contracts, options, futures, and many other financial derivatives are examples of contingent securities.

A {\em prediction market} is a securities market primarily focused on information aggregation.  For a future event with $n$ mutually exclusive and exhaustive possible outcomes, a typical prediction market offers $n$ Arrow-Debreu securities, each corresponding to a particular outcome. The prices of these securities form a probability distribution over the outcome space of the event, and can be viewed as the traders' collective estimate of the likelihood of each outcome. Market-based probability estimates have proved to be accurate in a variety of domains including business, entertainment, and politics~\cite{LHI09,Ber:01,Wol:04}.

Denote a set of mutually exclusive and exhaustive states of the world as $\O$. A securities market is \emph{complete} if there are $|\O| -1$ linearly independent securities~\cite{Arrow:64, Arrow:70, Mas:95}. For example, a prediction market with $n$ Arrow-Debreu securities for an $n$-outcome event is complete. With a complete securities market, any desired future payoff over the state space can be constructed by linearly combining these securities, which allows a trader to hedge any possible risk he may have.  Furthermore, traders can change the market prices to reflect any valid probability distribution over the state space, allowing them to reveal any information. Completeness therefore provides expressiveness for both risk allocation and information aggregation, making it a desirable property.  However, completeness is not always achievable.

In many real-world settings, the state space can be exponentially large or infinite. For instance, a competition among $n$ candidates results in a state space of $n!$ rank orders, while the future price of a stock has an infinite state space. In such situations, operating a complete securities market is not practical due to the notorious difficulties that humans have estimating small probabilities and the computational intractability of managing a large security set. It is natural to offer a smaller set of structured securities instead. For example, instead of having one security for each rank ordering, pair betting allows securities of the form ``\$1 if candidate A beats candidate B''. There has been a surge of recent research examining the tractability of running standard prediction market mechanisms (such as the popular Logarithmic Market Scoring Rule (LMSR) market maker~\cite{H03}) over combinatorial state spaces by limiting the space of available securities~\cite{PS07}. While this line of research has led to a few positive results~\cite{CGP08,Guo:09}, it has led more often to hardness results~\cite{CFLPW08} or to markets with undesirable properties such as unbounded loss of the market institution~\cite{GCP09}. 

In this paper, we propose a general framework to design {\em automated market makers} for securities markets. An automated market maker is a market institution who sets prices for each security and is always willing to accept trades at these prices. Unlike previous research aimed at finding a space of securities that can be efficiently priced using an existing market maker like LMSR, we start with an arbitrary space of securities and design a \emph{new} market maker tailored to this space. Our framework is therefore extremely general. Both LMSR and Quad-SCPM~\cite{ADPWY10} also fall under our framework.

We take an axiomatic approach. Given a space of securities with bounded payoff, we define a set of intuitive conditions that a reasonable market maker should satisfy. We prove that a market maker satisfying these conditions must price securities via a convex potential function, and that the space of reachable security prices must be precisely the convex hull of the security payoffs. We then incorporate ideas from online convex optimization~\cite{H09,R09} to define a convex cost function in terms of an optimization over the convex hull of the security payoffs. The optimal solution to the optimization problem gives the security prices. With this framework, we provide an efficient market for predicting the landing location of an object on a sphere.  

We then show that we can relax our ``no-arbitrage'' condition to design a new efficient market maker for pair betting, which is known to be \#P-hard to price using LMSR~\cite{CFLPW08}. This relaxation also allows the market maker to charge transaction fees so that the depth of the market can be dynamically increased as the number of trades increases, a desirable property that the extension of LMSR recently proposed by \citet{Othman:10b} was specifically designed to satisfy.

\paragraph{Preliminaries and Related Work:}

A simple cost function based market maker~\cite{H03,H07,CP07,CV10} offers $|\O|$ Arrow-Debreu securities, each corresponding to a potential outcome of an event.  The market maker determines how much each security should cost using a differentiable \emph{cost function}, $C: \reals^{|\O|} \rightarrow \reals$, which is simply a potential function specifying the amount of money currently wagered in the market as a function of the number of shares of each security that have been purchased.  If $q_{\o}$ is the number of shares of security $\o$ currently held by traders, and a trader would like to purchase a bundle of $r_{\o}$ shares for each security $\o \in \O$ (where some $r_\o$ could be zero or even negative, representing a sale), the trader must pay $C(\q + \r) - C(\q)$ to the market maker.  The instantaneous price of security $\o$ (that is, the price per share of an infinitely small portion of a security) is then $\partial C(\q)/\partial q_{\o}$, and is denoted $p_{\o}(\q)$.

The market designer is free to choose any differentiable cost function $C$ that satisfies a few basic properties.  First, it must be the case that for every $\o \in \O$ and every $\q \in \reals^{|\O|}$, $p_{\o}(\q) \geq 0$.  This ensures that the price of a security is never negative.  Second, if the market designer wishes to prevent arbitrage, it must be the case that for every $\q \in \reals^{|\O|}$, $\sum_{\o \in \O} p_{\o}(\q) =1$.  That is, the sum of the instantaneous prices of the securities must always be 1.  If the prices summed to something less than (respectively, greater than) 1, then a trader could purchase (respectively, sell) small equal quantities of each security for a guaranteed profit.\footnote{\citet{Othman:10b} recently analyzed a variation of LMSR in which $\sum_{\o \in \O} p_{\o}(\q) \geq 1$, violating this no-arbitrage condition.  We also explore relaxations of the no-arbitrage condition in Section~\ref{sec:relaxations}.}  These conditions ensure that the current prices can always be viewed as a probability distribution over the outcome space.  
One example of a cost function based market that has received considerable attention is Hanson's Logarithmic Market Scoring Rule (LMSR)~\cite{H03,H07,CP07}.  The cost function of the LMSR is $C(\q) = b \log \sum_{\o \in \O} \e^{q_{\o}/b}$, where $b > 0$ is a parameter of the market controlling the rate at which prices change.  The corresponding price function for each security $\o$ is $p_{\o}(\q) = \partial C(\q)/\partial q_{\o} = \e^{q_{\o}/b} / \sum_{\o' \in \O} \e^{q_{\o'}/b}$. 

%It is well known that the worst-case monetary loss of the LMSR is $b \log |\O|$. 

When $|\O|$ is large or infinite, calculating the cost of a purchase becomes intractable in general. Recent research has focused on restricting the allowable securities over a combinatorial outcome space and examining whether LMSR prices can be computed efficiently in the restricted space. If the outcome space contains $n!$ rank orders of $n$ competing candidates, it is \#P-hard for LMSR to price pair bets (e.g., ``\$1 if and only if candidate A beats candidate B'') or subset bets (e.g., ``\$1 if one of the candidates in subset $C$ finishes at position $k$'')~\cite{CFLPW08}. If the outcome space contains $2^n$ Boolean values of $n$ binary base events, it is \#P-hard for LMSR to price securities on conjunctions of any two base events (e.g., ``\$1 if and only if a Democrat wins Florida and Ohio'')~\cite{CFLPW08}. This line of research has led to some positive results when the uncertain event enforces particular structure on the outcome space. In particular, for a single-elimination tournament of $n$ teams, securities such as ``\$1 if and only if team A wins a $k$th round game'' and ``\$1 if and only if team A beats team B given they face off'' can be priced efficiently in LMSR~\cite{CGP08}. For a taxonomy tree on some statistic where the value of the statistic of a parent node is the sum of those of its children, securities such as ``\$1 if and only if the value of the statistic at node A belongs to $[x, y]$'' can be priced efficiently in LMSR~\cite{Guo:09}. 

Our paper takes a drastically different approach. Instead of searching for supportable spaces of securities for existing market makers, we design new market makers tailored to any security space of interest. Additionally, rather than requiring that securities have a fixed \$1 payoff when the underlying event happens, we allow more general contingent securities with arbitrary efficiently computable and bounded payoffs.

This research builds upon ideas from our earlier work~\cite{CV10} exploring the striking mathematical connections between complete cost function based prediction markets and no-regret learning.  In that work, we first showed that any complete cost function based prediction market can be interpreted as an algorithm for learning from expert advice by equating the set of outcomes or states of the world with the set of experts in the learning setting, and equating trades made in the market with expert losses.  Furthermore, we showed that if the loss of the market maker is bounded, this bound can be used to derive an $O(\sqrt{T})$ regret bound for the corresponding learning algorithm. That work focused entirely on complete markets, while here we explore how to use these connections to design new market making mechanisms for broader security spaces.

\ignore{
Othman~et~al.~\cite{Othman:10b} observed that security prices in LMSR only depend on the relative quantities purchased for every outcome. If the relative quantities do not change, even if the absolute quantities of all securities increase, that is the total amount of money invested in the market increases, the prices remain to be the same. This contradicts with the typical intuition of market depth -- when there are a lot of traders (i.e. a deep market), the security prices should change slowly. Othman~et~al. modified LMSR to allow the market maker to allow the market depth increase with purchases. Our section~\ref{sec:transactioncost} suggests a general approach of achieving adaptive market depth by imposing transaction costs on traders. 
}

\section{A New Framework for Market-Making Over Complex Security Spaces}

In the complete cost function based markets described above, the market maker offers an Arrow-Debreu security corresponding to each potential state of the world. We consider a market-design scenario where the state space $\O$ could potentially be quite large, or even infinite, making it infeasible to run such a market. Instead, we allow the market maker to offer a menu of $K$ securities for some reasonably-sized $K$, with the payoff of each security described by an arbitrary but efficiently-computable function $\rhob : \O \to \reals^K_+$.  Specifically, if a trader purchases a share of security $i$ and the outcome is $\o$, then the trader is paid $\rho_i(\o)$.  We call such security spaces \emph{complex}. A complex security space reduces to the complete security space if $K=|\O|$ and for each $i \in \{1,\cdots, K\}$, $\rho_i(\o) = 1$ if and only if $\o$ is the $i$th outcome.  We consider traders that purchase \emph{security bundles} $\r \in \reals^K$, and say that the payoff for $\r$ upon outcome $\o$ is exactly $\rhob(\o)\cdot\r$, where $\rhob(\o)$ denotes the vector of payoffs for each security for outcome $\o$. Let $\rhob(\O) = \{\rhob(\o)| \o \in \O\}$.

We do not presuppose a cost function based market.  However, in Section~\ref{sec:axioms}, we show that the use of a convex potential function is \emph{necessary} given some minor assumptions. In Section~\ref{sec:conjugatestuff}, we go on to show how to design an appropriate cost function by employing techniques from online convex optimization.

\subsection{Imposing Some Natural Restrictions on the Market Maker}
\label{sec:axioms}

In this section we introduce a sequence of conditions or axioms that one might expect a market to satisfy, and show that these conditions lead to some natural mathematical restrictions on the costs of security bundles. (We consider relaxations of these conditions in Section~\ref{sec:relaxations}.) Similar conditions were suggested for complete markets by \citet{CV10}, who defined the notion of a \emph{valid cost function}, and by \citet{Othman:10b}, who discussed properties similar to our notions of path independence and expressiveness, among others.

Imagine a sequence of traders entering the marketplace and purchasing security bundles. Let $\r_1, \r_2, \r_3, \ldots$ be the sequence of security bundles purchased. After $t-1$ such purchases, the $t$th trader should be able to enter the marketplace and query the market maker for the cost of arbitrary bundles. The market maker must be able to furnish a cost $\cost(\r | \r_1, \ldots, \r_{t-1})$ for any bundle $\r$. If the trader chooses to purchase $\r_t$ at a cost of $\cost(\r_t | \r_1, \ldots, \r_{t-1})$, the market maker may update the costs of each bundle accordingly.  Our first condition requires that the cost of acquiring a bundle $\r$ must be the same regardless of how the trader splits up the purchase.

\begin{condition}[Path Independence] \label{cond:pind}
For any $\r$, $\r'$, and $\r''$ such that $\r = \r' + \r''$, for any $\r_1, \ldots, \r_t$, 
$\cost(\r| \r_1, \ldots, \r_t) = \cost(\r'| \r_1, \ldots, \r_t) + \cost(\r''| \r_1, \ldots, \r_t, \r')$.
\end{condition}

It turns out that this condition \emph{alone} implies that prices can be represented by a cost function $C$, as illustrated in the following theorem.  The proof is by induction on $t$.\footnote{All omitted proofs appear in the appendix.}

\begin{theorem}
  Under Condition~\ref{cond:pind}, there exists a cost function $C : \reals^K \to \reals$ such that we may always write $\cost(\r_t | \r_1, \ldots, \r_{t-1}) = C(\r_1 + \ldots + \r_{t-1} + \r_t) - C(\r_1 + \ldots + \r_{t-1})$.  
\label{thm:needcostfunc}
\end{theorem}

With this in mind, we drop the cumbersome $\cost(\r | \r_1, \ldots, \r_{t})$ notation from now on, and write the cost of a bundle $\r$ as $C(\q + \r) - C(\q)$, where $\q = \r_1 + \ldots + \r_{t}$ is the vector of previous purchases.

Now, recall that one of the functions of a securities market is to aggregate traders' beliefs into an accurate prediction.  Each trader may have his own (potentially secret) information about the future, which we represent as a distribution $\p \in \simp_{|\O|}$ over the outcome space. The pricing mechanism should therefore incentivize the traders to reveal $\p$, while simultaneously avoid providing arbitrage opportunities.  Towards this goal, we introduce four additional conditions on our pricing mechanism.  

The first condition ensures that the gradient of $C$ is always well-defined. If we imagine that a trader can buy or sell an arbitrarily small bundle, we would like the cost of buying and selling an infinitesimally small quantity of any bundle to be the same.  If $\nabla C(\q)$ is well-defined, it can be interpreted as a vector of instantaneous prices for each security, with $\partial C(\q)/\partial q_{\o}$ representing the price per share of an infinitesimally small amount of security $\o$.
 Additionally, we can interpret $\nabla C(\q)$ as the traders' current estimates of the expected payoff of each security, in the same way that $\partial C(\q)/\partial q_{\o}$ was interpreted as the probability of outcome $\o$ when considering the complete security space.  

\ignore{
\yc{Let $\mathbf{e_i}$ be the unit vector with 1 for its $i$-th element and 0 everywhere else. The unit price of the purchase is $\frac{C(\q+r_i\mathbf{e_i}) - C(\q)}{r_i}$. If we require the unit price exists as $r_i$ approaches to 0 for any $i$, $r_i$, and $\q$, we get differentiability by definition. But I don't know whether this is too brutal.}
}

%\yc{Do we have a better way to define this?}
\begin{condition}[Existence of Instantaneous Prices]\label{cond:smooth}
 $C$ is continuous and differentiable everywhere.
\end{condition}

\ignore{
	The price of a tiny security bundles must eventually vanish. Furthermore, the pricing function must be smooth as a with respect to the previous security purchased. More precisely, for any $\r_1, \ldots, \r_t,\r,\r'$ we have that
	\begin{align*}
		\cost(\epsilon \r| \r_1, \ldots, \r_t) & \to 0 & \textnormal{ as } & \epsilon \to 0, \textnormal { and } \\
		\cost(\r'| \r_1, \ldots, \r_t + \epsilon \r) & \to \cost(\r'| \r_1, \ldots, \r_t) & \textnormal{ as } & \epsilon \to 0.
	\end{align*}
	\textbf{JDA: I notice now this last condition is wrong. We need a better way to motivate differentiability, which is what the latter is supposed to imply. Need to think about this...}
\end{condition}
}

The next condition encompasses the idea that the market should react to trades in a sensible way in order to incorporate the private information of the traders.  In particular, it says that the purchase of a security bundle $\r$ should never cause the market to lower the price of $\r$.  It turns out that this condition is closely related to incentive compatibility for a myopic trader. It is equivalent to requiring that a trader with a distribution  $\p \in \simp_{|\O|}$ can never find it simultaneously profitable (in expectation) to buy a bundle $\r$ or to buy the bundle $-\r$. In other words, there can not be more than one way to express one's information. 

\begin{condition}[Information Incorporation] \label{cond:info}
  For any $\q$ and $\r \in \reals^K$, $C(\q + 2\r) - C(\q + \r) \geq C(\q + \r) - C(\q)$.
\end{condition}

The no arbitrage condition states that it is never possible for a trader to purchase a security bundle $\r$ and receive a positive profit regardless of the outcome.  

\begin{condition}[No Arbitrage]\label{cond:arbfree}
 For all $\q, \r \in \reals^K$, there exists an $\o \in \O$ such that $C(\q + \r) - C(\q) \geq \r\cdot \rho(\o) $.  \end{condition}

Finally, the expressiveness condition specifies that any trader can set the market prices to reflect his beliefs about the expected payoffs of each security if  arbitrarily small portions of shares may be purchased. 

\begin{condition}[Expressiveness]\label{cond:express}
For any $\p \in \simp_{|\O|}$, $\exists \q \in \reals^K \cup \{\infty,-\infty\}$ for which $\nabla C(\q) = \E_{\o \sim \p} [\rhob(o)]$. 
\end{condition}

Let $\H(\cdot)$ denote a convex hull.  We characterize the form of the cost function under these conditions.

\begin{theorem}
	Under Conditions~\ref{cond:smooth}-\ref{cond:express}, $C$ must be convex with $\{\nabla C(\q) : \q \in \reals^K\} = \H(\rhob(\O))$. 
\label{thm:characterization}
\end{theorem}

Specifically, the existence of instantaneous prices implies that $\nabla C(\q)$ is well-defined.  The incorporation of information condition implies that $C$ is convex. The convexity of $C$ and the no arbitrage condition imply that $\{\nabla C(\q) : \q \in \reals^K\} \subseteq \H(\rhob(\O))$. Finally, the expressiveness condition is equivalent to requiring that $\H(\rhob(\O)) \subseteq \{\nabla C(\q) : \q \in \reals^K\}$.  

This theorem tells us that to satisfy our conditions, the set of reachable prices of a market should be {\em exactly} the convex hull of $\rhob(\O)$.  For complete markets, this would imply that the set of reachable prices should be precisely the set of all probability distributions over the $n$ outcomes.

\subsection{Designing the Cost Function via Conjugate Duality}
\label{sec:conjugatestuff}

The natural conditions we introduced above imply that to design a market for a set of $K$ securities with payoffs specified by an arbitrary payoff function $\rhob : \O \to \reals^K_+$, we should use a cost function based market with a convex, differentiable cost function such that $\{\nabla C(\q) : \q \in \reals^K\} = \H(\rhob(\O))$.  We now provide a general technique that can be used to design and compare properties of cost functions that satisfy these criteria.  In order to accomplish this, we make use of tools from convex analysis. 

It is well known\footnote{For a detailed discussion of convex conjugates and their properties, refer to a good text on convex optimization such as \citet{BV04} or \citet{HL04}.} that any closed, convex, differentiable function $C: \reals^K \rightarrow \reals$ can be written in the form $C(\q) = \sup_{\x \in \dom(R)} \x \cdot \q - R(\x)$ for a strictly convex function $R$ called the \emph{conjugate} of $C$.  (The strict convexity of $R$ follows from the differentiability of $C$.) Furthermore, any function that can be written in this form is convex.   As we will show in Section~\ref{sec:conjprops}, the gradient of $C$ can be expressed in terms of this conjugate: $\nabla C(\q) =\argmax_{\x \in \dom(R)} \x \cdot \q - R(\x)$.
To generate a convex cost function $C$ such that $\nabla C(\q) \in \Pi$ for all $\q$ for some set $\Pi$, it is therefore sufficient to choose an appropriate conjugate function $R$, restrict the domain of $R$ to $\Pi$, and define $C$ as 
\begin{equation}
C(\q) = \sup_{\x \in \Pi} \x \cdot \q - R(\x) ~.
\label{eqn:costfunc}
\end{equation}

We call such a market a \emph{complex cost function based market}.  To generate a cost function $C$ satisfying our five conditions, we need only to set $\Pi = \H(\rhob(\O))$ and select a strictly convex function $R$.  

This method of defining $C$ is convenient for several reasons.  First, it leads to markets that are efficient to implement whenever $\Pi$ can be described by a polynomial number of simple constraints.  Similar techniques have been applied to design learning algorithms in the online convex optimization framework~\cite{H09,R09}, where $R$ plays the role of a regularizer, and have been shown to be efficient in a variety of combinatorial applications, including online shortest paths, online learning of perfect matchings, and online cut set~\cite{CL10}.
Second, it yields simple formulas for properties of markets that help us choose the best market to run.  Two of these properties, worst-case monetary loss and worst-case information loss, are analyzed below.  

Note that both the LMSR and Quad-SCPM~\cite{ADPWY10} are examples of complex cost function based markets, though they are designed for the complete market setting only.

\subsection{Bounding Market Maker Loss and Loss of Information}
\label{sec:conjprops}

Before discussing market properties, it is useful to review some helpful properties of conjugates. The first is a convenient duality: For any convex, closed function $C$, the conjugate of the conjugate of $C$ is $C$ itself.  This implies that if $C$ is defined as in Equation~\ref{eqn:costfunc}, we may write
$R(\x) = \sup_{\q \in \reals^K} \q \cdot \x - C(\q)$.
Since this maximization is unconstrained, the maximum occurs when $\nabla C(\q) = \x$.  (Note that this may hold for many different values of $\q$.)  Suppose for a particular pair $(\x^*, \q^*)$ we have $\nabla C(\q^*) = \x^*$.  We can then rewrite this equation as
$R(\x^*)= \q^* \cdot \x^* -C(\q^*)$,
which gives us that 
$C(\q^*) = \q^* \cdot \x^* - R(\x^*)$.
From Equation~\ref{eqn:costfunc}, this tells us that $\x^*$ must be a maximizer of $\x \cdot \q - R(\x)$.  In fact, it is the unique maximizer due to strict convexity.
This implies, as mentioned above, that $\nabla C(\q) = \argmax_{\x \in \Pi} \x \cdot \q - R(\x)$.

By a similar argument we have that for any $\q$, if $\nabla R(\x) = \q$ then $\x$ maximizes $\x \cdot \q - R(\x)$ and therefore, as we have just shown, $\x = \nabla C(\q)$. However, the fact that $\x = \nabla C(\q)$ \emph{does not} imply that $\nabla R(\x) = \q$; in the markets we consider, it is generally the case that $\x = \nabla C(\q)$ for multiple $\q$.

\ignore{
 If $R$ is a barrier function (that is, if $\nabla R(\x) \rightarrow \infty$ as $\x$ approaches the border of $\dom(R)$), then the value of $\x$ maximizing $\x \cdot \q - R(\x)$ will always lie in the interior of $\dom(R)$.
}

We also make use of the notion of Bregman divergence.  The \emph{Bregman divergence} with respect to a convex function $f$ is given by $D_f(\x,\y) := f(\x) - f(\y) - \nabla f(\y)(\x - \y)$.  It is clear by convexity that $D_f(\x,\y) \geq 0$ for all $\x$ and $\y$.

\ignore{
Second, the divergence is invariant under linear perturbations of $f$. That is, if $f'(x) := f(\x) + \z \cdot \x + d$ for any vector $\z$ and constant $d$, we have that $D_f(\x,\y) = D_{f'}(\x,\y)$. Hence, if $\y$ minimizes $f$, then it's easy to see that $D_f(\x,\y) = f(\x) - f(\y)$. Finally, we have that
\begin{equation} \label{eq:scbound}
	D_f(\x,\y) \leq D_f(\x,\y)  + D_f(\y,\x)  = \langle \nabla f(\x) - \nabla f(\y), \x - \y \rangle \leq \|\nabla f(\x) - \nabla f(\y)\|^* \|\x - \y\| ~.
\end{equation}
The last inequality, which holds for any dual norm pair, follows by H\"{o}lder's inequality.  
}

\subsubsection{Bounding the Market Maker's Monetary Loss}

When comparing market mechanisms, it is useful to consider the market maker's worst-case monetary loss, 
$\sup_{\q \in \reals^K} 
\left( \sup_{\o \in \O} ( \rhob(\o) \cdot \q) - C(\q) + C(\0)\right)$.
This quantity is simply the worst-case difference between the maximum amount that the market maker might have to pay the traders ($\sup_{\o \in \O}  \rhob(\o) \cdot \q$) and the amount of money collected by the market maker ($C(\q) - C(\0)$).  The following theorem provides a bound on this loss in terms of the conjugate function $R$.

\begin{theorem} 
Consider any complex cost function based market with $\Pi = \H(\rhob(\O))$.  Let $\q$ denote the vector of quantities sold and $\o$ denote the true outcome.  The monetary loss of the market maker is no more than
\[
R(\rhob(\o)) - \min_{\x \in \H(\rhob(\O))} R(\x) - D_R(\rhob(\o), \nabla C(\q))~.
\]
Consequently, the worst-case market maker loss is no more than $\sup_{\x \in \rhob(\O)} R(\x) - \min_{\x \in \H(\rhob(\O))} R(\x)$.
\label{thm:worstcaseloss}
\end{theorem}

This theorem tells us that as long as the conjugate function is bounded on $\H(\rhob(\O))$, the market maker's worst-case loss is also bounded.  Furthermore, it quantifies the intuitive notion that the market maker will have higher profits when the distance between $\rhob(\o)$ and the final vector of prices $\nabla C(\q)$ is large.  Viewed another way, the market maker will pay more when $\nabla C(\q)$ is a good estimate of $\rhob(\o)$.

\ignore{
\begin{proof}
We can bound the worst-case monetary loss as
\begin{eqnarray*}
\lefteqn{\sup_{\q \in \reals^K} \left( \max_{i \in \{1,\cdots,N\}} \m_i \cdot \q -  (C(\q) - C(\0)) \right) }
\\
&\leq& \sup_{\q \in \reals^K} \left( \sup_{\x \in \HM} \x \cdot \q - R(\x) + R(\x) - C(\q) \right) + C(\0)
\leq \sup_{\q \in \reals^K} R(\x) + C(\0)
\\
&=& \sup_{\q \in \reals^K} R(\x)  - \inf_{\q \in \reals^K} R(\x) ~.
\end{eqnarray*}
\end{proof}
} % end ignore of proof

\subsubsection{Bounding Information Loss} \label{sec:infoloss}

Information loss can occur when securities are sold in discrete quantities (for example, single units), as they are in most real-world markets.  Without the ability to purchase arbitrarily small bundles, traders may not be able to change the market prices to reflect their true beliefs about the expected payoff of each security, even if expressiveness is satisfied.  We will argue that the amount of information lost is captured by the market's bid-ask spread for the smallest trading unit.  Given some $\q$, the current bid-ask spread of security bundle $\r$ is defined to be $\left(C(\q + \r) - C(\q)\right) - \left(C(\q) - C(\q - \r)\right)$.  This is simply the difference between the current cost of buying the bundle $\r$ and the current price at which $\r$ could be sold.

To see how the bid-ask spread relates to information loss, suppose that the current vector of quantities sold is $\q$.  If securities must be sold in unit chunks, a rational, risk-neutral trader will not buy security $i$ unless she believes the expected payoff of this security is at least $C(\q + \eb_i) - C(\q)$.  Similarly, she will not sell security $i$ unless she believes the expected payoff is at most $C(\q) - C(\q - \eb_i)$.  If her estimate of the expected payoff of the security is between these two values, she has no incentive to buy or sell the security.  In this case,  it is only possible to infer that the trader believes the true expected payoff lies somewhere in the range $[C(\q) - C(\q - \eb_i), C(\q + \eb_i) - C(\q)]$.  The bid-ask spread is precisely the size of this range.

Intuitively, the bid-ask spread relates to the depth of the market. When the bid-ask spread is small, a small order can change the prices of the securities dramatically. The market is shallow. When the bid-ask spread is large, large orders may only move the prices slightly. The market is deep. The bid-ask spread depends on how fast the instantaneous prices change. For complete markets, Chen and Pennock~\cite{CP07} use the inverse of $\partial^2 C(\q)/\partial q_{\o}^2$ to capture this notion for each security $\o$ independently. We define a  \emph{market depth parameter}, $\beta$, for our complex securities markets with twice-differentiable $C$ in a similar spirit. We will bound the bid-ask spread in terms of this parameter. Using the market depth parameter, it is easy to see that there exists a clear trade-off between worst-case monetary loss and information loss.   

\begin{definition}
For any complex cost function based market, if $C$ is twice-differentiable, the {\em market depth parameter} $\beta(\q)$ for a quantity vector $\q$ is defined as $\beta(\q) = 1/V_c(\q)$, where $V_c(\q)$ is the largest eigenvalue of $\nabla^2C(\q)$, the Hessian of $C$ at $\q$. The worst-case market depth is $\beta = \inf_{\q \in \reals^K} \beta(\q)$.  
\end{definition}

% We note that $C$ being twice differentiable is equivalent to $C$ having a twice-differentiable $R$. 
Let $\relint(\Pi)$ be the relative interior of $\Pi$. If $C$ is twice-differentiable, then for any $\q$ such that $\nabla C(\q) \in \relint(\Pi)$, we have a correspondence between the Hessian of $C$ at $\q$ and the Hessian of $R$ at $\nabla C(\q)$. More precisely, we have that
%\begin{equation}\label{eq:conj_hess}
$\u^\top\nabla^2 C(\q) \u = \u^\top\nabla^{-2} R(\nabla C(\q)) \u$ for any $\u = \x - \x'$ with $\x,\x' \in \Pi$. (See, for example, Gorni \cite{Gorni1991293} for more.) This means that $\beta(\q)$ is equivalently defined as the smallest eigenvalue of $\nabla^{2} R(\nabla C(\q)) \arrowvert_{\Pi}$; that is, where we consider the second derivative only within the price region $\Pi$.  

The definition of worst-case market depth implies that  $1/\beta$ is an upper bound on the curvature of $C$, which implies that $C$ is locally bounded by a quadratic with Hessian $I/\beta$. We can derive the following.

%end ignore

\begin{lemma}
	Consider a complex cost function based market with worst-case market depth $\beta$. For any $\q$ and $\r$ we have $D_C(\q + \r, \q) \leq   \|\r\|^2 / (2\beta)$. 
\label{lem:Dcbound}
\end{lemma}

It is easy to verify that the bid-ask spread can be written in terms of Bregman divergences.  In particular, $C(\q + \r) - C(\q) - \left(C(\q) - C(\q - \r)\right) = D_C(\q + \r, \q) + D_C(\q - \r, \q)$.  This implies that the worst-case bid-ask spread of a market with market depth $\beta$ can be upperbounded by a constant times $1/\beta$.  That is, as the market depth parameter increases, the bid-ask spread must decrease.  The following theorem shows that this leads to an inherent tension between worst-case monetary loss and information loss.

\begin{theorem}
	For any complex cost function based market with worst-case market depth $\beta$, for any $\r$, $\q$ meeting the conditions in Lemma~\ref{lem:Dcbound}, the bid-ask spread for bundle $\r$ with previous purchases $\q$ is no more than $2 \|\r\|^2 / \beta$.  The worst-case monetary loss of the market maker at least $\beta \cdot  \textnormal{diam}^2(\H(\rhob(\O)))/8$. 
\label{thm:betatheorem}
\end{theorem}

% We shall now define a notion of \emph{liquidity} of a market, which shall measure how slowly the prices change for security purchases. That is, a high liquidity will be equivalent to a uniformly small buy/sell spread. For a moment, let's imagine that $C$ is twice differentiable. If we consider a small security bundle $\r$, then one side of the buy/sell spread is
% \[
% 	D_C(\q + \r, \q) \approx \r^\top \nabla^2 C(\q) \r.
% \]
% As we would like to scale inversely with the buy/sell spread, and we would like to define it independently of $\r$, a natural choice for the liquidity might be the inverse of the largest eigenvalue of $\nabla^2 C(\q)$. Unfortunately, this may be undefined when $C$ is not twice-differentiable, so we begin with a more general definition.

We can see that there is a direct trade-off between the upper bound of the bid-ask spread, which shrinks as $\beta$ grows, and the lower bound of the worst-case loss of the market maker, which grows linearly in $\beta$. This trade-off is very intuitive.  When the market is shallow (small $\beta$), small trades have a large impact on market prices, and traders cannot purchase too many shares of the same security without paying a lot.  When the market is deep (large $\beta$), prices change slowly, allowing the market maker to gain more precise information, but simultaneously forcing the market maker to take on more risk since many shares of a security can be purchased at prices that are potentially too low.  This trade-off can be adjusted by scaling $R$, which scales $\beta$.  This is analogous to adjusting the ``liquidity parameter'' $b$ in the LMSR.

% The following theorem bounds the bid-ask spread (and therefore the information loss of the market) in terms of the divergence.
% 
% \jenn{We would like to bound this in terms of $R$ instead of $C$.  Jake said he had ideas about how to do this.}
% 
% \begin{theorem}
% Consider a market maker selling a set of $K$ securities with payoffs defined by a function $\rhob : \O \to \reals^K_+$ using the cost function specified in Equation~\ref{eqn:costfunc}.  Let $\q$ denote the vector of quantities sold.  The current bid-ask spread for the $i$th security is exactly
% \[
% D_C(\q + \eb_i, \q) + D_C(\q - \eb_i, \q) ~.
% \]
% \label{thm:buysellspread}
% \end{theorem}
% 

\subsection{An Example}

To illustrate the use of our framework for market design, we consider the following example. An object orbiting the planet, perhaps a satellite, is predicted to fall to earth in the near future and will land at an unknown location, which we would like to predict.  We represent locations on the earth as unit vectors $\u \in \reals^3$. We will design a market with three securities, each corresponding to one coordinate of the final location of the object. In particular, security $i$ will pay off $u_i + 1$ dollars if the object lands in location $\u$. (The addition of $1$, while not strictly necessary, ensures that the payoffs, and therefore prices, remain positive, though it will be necessary for traders to sell securities to express certain beliefs.)  This means that traders can purchase security bundles $\r \in \reals^3$ and, when the object lands at a location $\u$, receive a payoff $(\u + \1) \cdot \r$.  Note that in this example, the outcome space is infinite, but the security space is small.

The price space $\H(\rhob(\O))$ for this market will be the 2-norm unit ball centered at $\1$. To construct a market for this scenario, let us make the simple choice of $R(\x) = \lambda \| \x - \1 \|^2$ for some parameter $\lambda > 0$. 
When $\|\q\| \leq 2 \lambda$, there exists an $\x$ such that $\nabla R(\x) = \q$.  In particular, this is true for $\x = (1/2) \q / \lambda + \1$, and $\q \cdot \x - R(\x)$ is minimized at this point.  When $\|\q\| > 2 \lambda$, $\q \cdot \x - R(\x)$ is minimized at an $\x$ on the boundary of $\H(\rhob(\O))$. Specifically, it is minimized at $\x = \q/||\q|| + \1$.
From this, we can compute
\[
	C(\q) = \begin{cases}
		\frac 1 {4\lambda} \|\q\|^2 + \q\cdot \1, &\text{ when } \|\q\| \leq 2 \lambda,\\
		\| \q \| + \q \cdot \1 - \lambda, &\text{ when } \|\q\| > 2 \lambda.
	\end{cases}
\]
The market depth parameter $\beta$ is $2\lambda$; in fact, $\beta(\x) = 2\lambda$ for any price vector $\x$ in the interior of $\H(\rhob(\O))$. By Theorem~\ref{thm:worstcaseloss}, the worst-case loss of the market maker is no more than $\lambda$, which is precisely the lower bound implied by Theorem~\ref{thm:betatheorem}. Finally, the divergence $D_C(\q + \r, \q) \leq \|\r\|^2 / (4\lambda)$ for all $\q,\r$, with equality when $\|\q\|,\|\q + \r\| \leq 2\lambda$, implying that the bid-ask spread scales linearly with $\|\r\|^2 / \lambda$.

\section{Relaxing the Feasible Price Region}
\label{sec:relaxations}

Thus far, we have argued that the space of feasible price vectors should be precisely $\H(\rhob(\O))$. In this section, we consider more general price spaces. This generalization can be beneficial when dealing with security spaces for which the convex hull $\H(\rhob(\O))$ is difficult to optimize over directly.  It also enables the design of markets that grow increasingly deep without requiring a sacrifice in terms of worst-case loss.

As before, each market will be defined in terms of a pair $(\Pi,R)$ where $\Pi \subseteq \reals^d$ is a convex compact set of feasible prices and $R: \reals^d \to \reals$ is a strictly convex function with domain $\Pi$. The market's cost function $C$ will be the conjugate of $R$ with respect to the set $\Pi$, as in Equation~\ref{eqn:costfunc}.  The only difference is that we now allow $\Pi$ to be distinct from $\H(\rhob(\O))$. Not surprisingly, the choice of $\Pi$ will affect the interest of the traders and the market maker. We prove several claims which will aid us in our market design. Proposition~\ref{prop:unbounded} tells us that the expressiveness condition should not be relaxed, while Proposition~\ref{prop:newworstcaseloss} tells us that the no-arbitrage condition can be.  Together, these imply that we may safely choose $\Pi$ to be a \emph{superset} of $\H(\rhob(\O))$.

\begin{proposition}
For any complex cost function based market, the worst-case loss of the market maker is unbounded if $\rhob(\O) \nsubseteq \Pi$. 
\label{prop:unbounded}
\end{proposition}

This (perhaps surprising) proposition tells us that expressiveness is not only useful for information aggregation, it is actually \emph{necessary} for the market maker to avoid unbounded loss. The proof involves showing that if $\o$ is the final outcome and $\rhob(\o) \not\in \Pi$, then it is possible to make an infinite sequence of trades such that each trade causes a constant amount of loss to the market maker. 

In the following proposition, which is a simple extension of Theorem~\ref{thm:worstcaseloss}, we see that including additional price vectors in $\Pi$ does not adversely impact the market maker's worst-case loss, despite the fact that the no-arbitrage condition is violated.

\begin{proposition}\label{prop:wcl_relax}
Consider any complex cost function based market with $R$ and $\Pi$ satisfying $\sup_{\x \in \H(\rhob(\O))} R(\x) < \infty$ and $\H(\rhob(\O)) \subseteq \Pi$. Assume that the initial price vector satisfies $\nabla C(\0) \in \H(\rhob(\O))$. Let $\q$ denote the vector of quantities sold and $\o$ denote the true outcome.  The monetary loss of the market maker is no more than
$R(\rhob(\o))  - \min_{\x \in \H(\rhob(\O))} R(\x) - D_R(\rhob(\o), \nabla C(\q)) $.
\label{prop:newworstcaseloss}
\end{proposition}

This  tells us that expanding $\Pi$ can only help the market maker; increasing the range of $\nabla C(\q)$ can only increase the divergence term. This may seem somewhat counterintuitive. We originally required that $\Pi \subseteq \H(\rhob(\O))$ as a consequence of the no-arbitrage condition, and by relaxing this condition, we are providing traders with potential arbitrage opportunities. However, these arbitrage opportunities do not hurt the market maker. As long as the initial price vector lies in $\H(\rhob(\O))$, any such situations where a trader can earn a guaranteed profit are effectively created (and paid for) by other traders!  In fact, if the final price vector $\nabla C(\q)$ falls outside the convex hull, the divergence term will be strictly positive, improving the bound.

To elaborate on this point, let's consider an example where $\Pi$ is strictly larger than $\H(\rhob(\O))$. Let $\q$ be the current vector of purchases, and assume the associated price vector $\x = \nabla C(\q)$ lies in the interior of $\H(\rhob(\O))$. Consider a trader who purchases a bundle $\r$ such that the new price vector leaves this set, i.e., $\y := \nabla C(\q + \r) \notin \H(\rhob(\O))$. We claim that this choice can be strictly improved in the sense that there is an alternative bundle $\r'$ whose associated profit, for any outcome $\o$, is strictly greater than the profit for $\r$.

% JAKE: I didn't really mean to include this fig in the final version, just to show you two
% \begin{figure}[htbp]
% 	\centering
% 		\includegraphics[height=2in]{ink.jpg}
% 	\label{fig:ink}
% \end{figure}

For simplicity, assume $\y$ is an interior point of $\Pi \setminus \H(\rhob(\O))$ so that $\q + \r = \nabla R(\y)$. Define $\pi(\y) := \arg\min_{\y' \in \H(\rhob(\O))} D_R(\y',\y)$, the minimum divergence projection of $\y$ into $\H(\rhob(\O))$. The alternative bundle we consider is $\r' = \nabla R(\pi(\y)) - \q$. Our trader pays $C(\q + \r) - C(\q + \r')$ less to purchase $\r'$ than to purchase $\r$. Hence, for any outcome $\o$, we see that the increased profit for $\r'$ over $\r$ is 
\begin{eqnarray} 
	\nonumber \rhob(\o) \cdot (\r' - \r) - C(\q + \r') + C(\q + \r) & > &  \rhob(\o)\cdot (\r' - \r) + \nabla C(\q + \r') \cdot (\r - \r') \\ 
	& = & (\rhob(\o) - \pi(\y)) \cdot (\r' - \r). \label{eq:profinc}
\end{eqnarray}
Notice that we achieve strict inequality precisely because $\nabla C(\q + \r') \ne \nabla C(\q + \r)$.
Now use the optimality condition for $\pi(\y)$ to see that, since $\rhob(\o) \in \H(\rhob(\O))$,
$\nabla_{\pi(\y)}(D_R(\pi(\y),\y)) \cdot (\rhob(\o) - \pi(\y)) \geq 0$.
It is easy to check that $\nabla_{\pi(\y)}(D_R(\pi(\y),\y)) = \nabla R(\pi(\y)) - \nabla R(\y) = \r' - \r$. Combining this last expression with the inequality above and \eqref{eq:profinc} tells us that the profit increase is strictly greater than
$(\rhob(\o) - \pi(\y)) \cdot (\r' - \r) \geq 0$.
Simply put, the trader receives a guaranteed positive increase in profit for any outcome $\o$.

The next proposition shows that any time the price vector lies outside of $\rhob(\o)$, traders could profit by moving it back inside. The proof uses a nice application of minimax duality for convex-concave functions.
\begin{proposition}
	For any complex cost function based market, given a current quantity vector $\q_0$ with current price vector $\nabla C(\q_0) = \x_0$, a trader has the opportunity to earn a guaranteed profit of at least $\min_{\x \in \H(\rhob(\O))} D_R(\x,\x_0)$.
\label{prop:arbitragebound}
\end{proposition}
In the case that $\x_0 \in \H(\rhob(\O))$, $D_R(\x,\x_0)$ is minimized when $\x = \x_0$ and the bound is vacuous, as we would expect. The more interesting case occurs when the prices have fallen outside of $\H(\rhob(\O))$, in which case a trader is guaranteed a riskless profit by moving the price vector to the closest point in $\H(\rhob(\O))$.

\subsection{Pair-betting Market via Relaxation}

We now turn our attention to the design of a prediction market for the scenario in which the outcome is some \emph{ranking} of a set of $n$ competitors, such as $n$ horses in a race. The outcome of such a race is a permutation $\pi : [n] \to [n]$, where $\pi(i)$ is the final position of $i$, with $\pi(i) = 1$ being best, and $[n]$ denotes the set $\{1,\cdots,n\}$.  A typical market for this setting might offer $n$ Arrow-Debreu securities, with the $i$th security paying off if and only if $\pi(i) = 1$.  Additionally, there might be a separate, \emph{independent} markets allowing bets on horses to place (come in first or second) or show (come in first, second, or third).  However, running independent markets for sets of outcomes with clear correlations is wasteful in that information revealed in one market does not automatically propagate to the others.  Instead, we consider a complex market in which bettors can make arbitrary \emph{pair bets}: for every $i,j$, a bettor can purchase a security which pays out if and only if $\pi(i) > \pi(j)$.  Pricing such bets using LMSR is known to be \#P-hard~\cite{CFLPW08}.

We can represent the payoff structure of any such outcome $\pi$ by a matrix $M_\pi$ defined by
\[
	M_\pi(i,j) = \begin{cases}
		1, &\text{ if }\pi(i) > \pi(j)\\
		\frac 1 2 , &\text{ if } i = j\\
		0, &\text{ if }\pi(i) < \pi(j).
	\end{cases}
\]
We would like to choose our feasible price region as the set $\H(\{ M_\pi : \pi \in S_n \})$, where $S_n$ is the set of permutations on $[n]$. Unfortunately, the computation of this convex hull is necessarily hard: if given only a separation oracle for the set $\H(\{ M_\pi : \pi \in S_n \})$, we could construct a linear program to solve the ``minimum feedback arcset'' problem, which is known to be NP-hard. 

On the positive side, we see from the previous section that the market maker can work in a larger feasible price space without risking a larger loss. We thus relax our feasible price region $\Pi$ to the set of matrices $X \in \reals^{n^2}$ satisfying
\begin{align*}
	X(i,j) & \geq 0 & \forall i,j \in [n] \\ 
	X(i,j) & = 1 - X(j,i) & \forall i,j \in [n] \\ 
	X(i,j) + X(j,k) + X(k,i) & \geq 1 & \forall i,j,k \in [n] 
\end{align*}

This relaxation was first discussed by Meggido \cite{megiddo1977mixtures}, who referred to such matrices as \emph{generalized order matrices}. He proved that, for $n \leq 4$, we do have $\Pi = \H(\{ M_\pi : \pi \in S_n \})$, but gave a counterexample showing strict containment for $n = 13$. By using this relaxed price space, the market maker allows traders to bring the price vector outside of the convex hull, yet includes a set of basic (and natural) constraints on the prices. 
Such a market could be implemented with any strongly convex conjugate function (e.g., quadratic).

\subsection{Transaction Costs via Relaxation}
\label{sec:transactioncost}

Let us return our attention to Proposition~\ref{prop:wcl_relax}, which bounds the worst-case loss of the market maker. Notice the term $R(\rhob(\o))  - \min_{\x \in \Pi} R(\x)$ is strictly positive, and accounts for the market maker's risk in offering the market. On the other hand, the term $ -D_R(\rhob(\o), \nabla C(\q))$ is non-positive, representing the potential profit that the market maker can earn if the final price vector $\nabla C(\q)$ is far from $\rhob(\o)$. The potential to make a profit may be appealing, but note that this term will approach $0$ if $\nabla C(\q)$ approaches $\rhob(\o)$ as traders gain information.  (Consider the behavior of traders in an election market as votes start to be tallied.)
%
%On the downside, it's quite possible that the traders gain increasing information so that $\nabla C(\q)$ approaches $\rhob(\o)$, hence the market can not guarantee a profit.
%
As discussed in Section~\ref{sec:infoloss}, the market maker can reduce his worst-case loss by adjusting the depth parameter, but this will result in a shallow market with a larger bid-ask spread.

To combat these problems, we propose a technique that allows us to guarantee lower worst-case loss without creating a shallow market by relaxing the feasible price region $\Pi$.  This relaxation is akin to introducing a transaction cost. By constructing the correct conjugate function, we may obtain a market with a bounded worst-case loss, a potential profit, and market depth that grows with the number of trades.

For simplicity, we consider the complete market scenario in which the marker maker offers an Arrow-Debreu security for every outcome. Let $\o_i$ denote outcome $i$ and $|\O|=n$. In this case, $\rhob(\o_i) = \eb_i$, and $\H(\rhob(\O)) = \Delta_n$, the $n$-simplex. We define $\Pi := \{ \x \in \reals^n : x_i \geq 0\; \forall i, 1 \leq \sum_{j=1}^n x_j \leq 1 + c\}$, where $c$ is some maximal transaction cost. The transaction cost is not imposed on individual traders or individual securities, but is split among all securities.
We also introduce the requirement that traders can only purchase \emph{positive} bundles $\r \in \reals_+^n$. In general, this restriction could prevent the traders from expressing ``negative'' beliefs, as we are disallowing the explicit shorting of securities, but in this particular market traders still have the ability to effectively short a security by purchasing equal quantities of the other $n-1$ securities. 

We must now choose any conjugate function $R$ satisfying the the following conditions:
\begin{enumerate}
	\item $R$ grows no larger than a constant within $\Delta_n$, so the worst-case loss $R(\eb_i) - R(\x_0)$ remains small.
	\item Outside of $\Delta_n$, $R(\x)$ becomes increasingly curved as $\x$ approaches the constraint $\sum_{j=1}^n x_j = 1 + c$. (Notice that the price vector $\x$ is guaranteed to approach this constraint as purchases are made since we allow only positive bundles to be purchased and thus $\q$ can only grow.)  Hence, the smallest eigenvalue of $\nabla^2 R(\x)$, which is the market depth at $\x$, must grow large as $\x$ approaches this boundary.
\end{enumerate}

The construction we have proposed here has several nice properties. It has bounded worst-case loss due to condition 1, and increasing market depth by condition 2. It imposes a transaction cost by letting the prices leave the simplex, but it does so in a smooth fashion; the sum of prices only approaches the value $1+c$ after many trades have occurred, and at this point the market depth will have become large. Lastly, as a result of condition 2, we know that the market maker's earnings increase as more trades occur, since $\min_{\x' \in \H(\rhob(\O))} D_R(\x',\x)$ must eventually increase
as the price vector $\x$ approaches the $1+c$ constraint.

The idea of introducing transaction costs to allow increasing market depth was also proposed by Othman~et~al.~\cite{Othman:10b}, who introduced a modified LMSR market maker with a particular cost function $C$. Their market can be viewed as a special case of our approach, although it is not defined via conjugate duality. In particular, they set the feasible price region of their market maker as a convex subset of $\{ \x \in \reals^n : x_i \geq 0\; \forall i, 1 \leq \sum_{j=1}^n x_j \leq 1 + \alpha n \log n\}$ for a positive parameter $\alpha$.

\newpage

{\small
\bibliographystyle{plainnat}
{\bibliography{marketsml}}
}

\section*{Appendix}

Below we provide all proofs that were omitted from the paper.

\subsection{Proof of Theorem~\ref{thm:needcostfunc}}

Let $C(\q) := \cost(\q | \emptyset)$. We shall prove, via induction, that for any $t$ and any bundle sequence $\r_1, \ldots, \r_{t}$,
	\begin{equation}\label{eq:cbased_ind}
		\cost(\r_t | \r_1, \ldots, \r_{t-1}) = C(\r_1 + \ldots + \r_{t-1} + \r_t) - C(\r_1 +  \ldots + \r_{t-1}) ~.
	\end{equation}
	
	When $t = 1$, this holds trivially. Assume that Equation~\ref{eq:cbased_ind} holds for all bundle sequences of any length $t \leq T$. By Condition~\ref{cond:pind},
	\begin{eqnarray*}
	\lefteqn{\cost(\r_{T+1} | \r_1, \ldots, \r_{T}) }\\
		& = & \cost(\r_{T+1} + \r_{T}| \r_1, \ldots, \r_{T-1}) - \cost(\r_{T}| \r_1, \ldots, \r_{T-1})\\
		& = & C\left(\r_{T+1} + \r_{T} + \sum_{t=1}^{T-1} \r_t \right) - C\left(\sum_{t=1}^{T-1} \r_t \right) - \left(C\left(\r_{T} + \sum_{t=1}^{T-1} \r_t \right) - C\left(\sum_{t=1}^{T-1} \r_t \right)\right)\\ 
		& = & C\left(\sum_{t=1}^{T+1} \r_t \right) - C\left(\sum_{t=1}^{T} \r_t \right) ~,
	\end{eqnarray*}
and we see that Equation~\ref{eq:cbased_ind} holds for $t=T+1$ too.
\qed

\subsection{Proof of Theorem~\ref{thm:characterization}}

We first prove convexity.  Assume $C$ is non-convex somewhere. Then there must exist some $\q$ and $\r$ such that $C(\q) > (1/2) C(\q + \r) + (1/2) C(\q - \r)$.  This means $C(\q+\r) - C(\q) < C(\q) - C(\q - \r)$, which contradicts Condition~\ref{cond:info}, so $C$ must be convex.

Now, Condition~\ref{cond:smooth} trivially guarantees that $\nabla C(\q)$ is well-defined for any $\q$. 
To see that $\{\nabla C(\q) : \q \in \reals^K\} \subseteq \H(\rhob(\O))$, let us assume there exists some $\q'$ for which $\nabla C(\q') \notin \H(\rhob(\O))$. This can be reformulated in the following way: There must exists some halfspace, defined by a normal vector $\r$, that separates $\nabla C(\q')$ from every member of $\rhob(\O)$. More precisely
	\[
		\nabla C(\q') \notin \H(\rhob(\O)) \quad \Longleftrightarrow \quad \exists \r\, \forall \o \in \O: \quad \nabla C(\q') \cdot \r \leq \rhob(\o) \cdot \r.
	\]
	On the other hand, letting $\q := \q' - \r$, we see by convexity of $C$ that $C(\q + \r) - C(\q) \leq \nabla C(\q')\cdot \r$. Combining these last two inequalities, we see that the price of bundle $\r$ purchased with history $\q$ is always smaller than the payoff for \emph{any} outcome. This implies that there exists some arbitrage opportunity, contradicting Condition~\ref{cond:arbfree}.

Finally, since $\H(\rhob(\O)=\{\E_{\o \sim \p} [\rhob(o)]| \p \in \simp_{|\O|}\}$, Condition~\ref{cond:express} implies that $\H(\rhob(\O)) \subseteq \{\nabla C(\q) : \q \in \reals^K\}$. 
\qed

\subsection{Proof of Theorem~\ref{thm:worstcaseloss}}

Since $\q$ is the final quantity vector, $\nabla C(\q)$ is the final vector of instantaneous prices.  From Equation~\ref{eqn:costfunc}, we have that $C(\q) = \nabla C(\q) \cdot \q - R(\nabla C(\q)) $ and  $C(\0) = -\min_{\x \in \H(\rhob(\O))} R(\x)$.   The difference between the amount that the market maker must pay out and the amount that the market maker has previously collected is then 
\begin{eqnarray*}
\lefteqn{\rhob(\o) \cdot \q - C(\q) + C(\0)}
\\
&=& \rhob(\o) \cdot \q - \left(\nabla C(\q) \cdot \q - R(\nabla C(\q))\right) - \min_{\x \in \H(\rhob(\O))} R(\x)
\\
&=& \q \cdot (\rhob(\o) - \nabla C(\q)) + R(\nabla C(\q)) - \min_{\x \in \H(\rhob(\O))} R(\x) + R(\rhob(\o)) - R(\rhob(\o))
\\
&=&  R(\rhob(\o)) - \min_{\x \in \H(\rhob(\O))} R(\x)
- \left(R(\rhob(\o)) - R(\nabla C(\q)) -  \q \cdot (\rhob(\o) - \nabla C(\q)) \right)
\\
&\leq& R(\rhob(\o)) - \min_{\x \in \H(\rhob(\O))} R(\x)
- \left(R(\rhob(\o)) - R(\nabla C(\q)) -  \nabla R(\nabla C(\q)) \cdot (\rhob(\o) - \nabla C(\q)) \right)
\\
&=& R(\rhob(\o)) - \min_{\x \in \H(\rhob(\O))} R(\x) - D_R(\rhob(\o), \nabla C(\q))~,
\end{eqnarray*}
where $D_R$ is the Bregman divergence with respect to $R$, as defined above.  The inequality follows from the first-order optimality condition for convex optimization:
\[
		\x = \arg\min_{\x' \in \Pi} f(\x') \quad \implies \quad \nabla f(\x) \cdot (\y - \x) \geq 0 \;\; \text{ for any } \y \in \Pi.
\]

Since the divergence is always nonnegative, this is upperbounded by $R(\rhob(\o)) - \min_{\x \in \H(\rhob(\O))} R(\x)$, which is in turn upperbounded by $\sup_{\x \in \rhob(\O)} R(\x) - \min_{\x \in \H(\rhob(\O))} R(\x)$.
\qed

% Cut this theorem
\ignore{
\subsection{Proof of Theorem~\ref{thm:buysellspread}}
\begin{eqnarray*}
\lefteqn{\left(C(\q + \eb_i) - C(\q)\right) - \left(C(\q) - C(\q - \eb_i)\right)}
\\
&=& C(\q + \eb_i) - C(\q) - \nabla C(\q) \cdot \eb_i
+ C(\q - \eb_i) - C(\q) - \nabla C(\q) \cdot (- \eb_i)
\\
&=&
D_C(\q + \eb_i, \q) + D_C(\q - \eb_i, \q) ~.
\end{eqnarray*}
\qed
}

\subsection{Proof of Theorem~\ref{thm:betatheorem}}

  The bound on the bid-ask spread follows immediately from Lemma~\ref{lem:Dcbound} and the argument above. The value $\beta$ lower-bounds the eigenvalues of $R$ everywhere on $\Pi$. Hence, if we do a quadratic lower-bound of $R$ from the point $\x_0 = \arg\min_{\x \in \Pi}R(\x)$ with Hessian defined by $\beta I$, then we see that $R(\x) - R(\x_0) \geq D_R(\x,\x_0) \geq \frac \beta 2  \|\x - \x_0\|^2$. In the worst-case, $\|\x - \x_0\| = \textnormal{diam}(\H(\rhob(\O)))/2$, which finishes the proof.
\qed

\subsection{Proof of Proposition~\ref{prop:unbounded}}

Consider some outcome $\o$ such that $\rho(\o) \notin \Pi$. 	
The feasible price set $\Pi=\{ \nabla C(\q) : \forall \q \}$ is compact. Because $\rho(\o) \notin \Pi$, there exists a hyperplane that {\em strongly} separates $\Pi$ and $\rho(\o)$. In other words, there exists an $k >0$ such that $||\rho(\o) - \nabla C(\q)|| \geq k$. 

When outcome $\o$ is realized, $B(\q) = \rhob(\o)\cdot \q - C(\q) + C(\0)$ is the market maker's loss given $\q$. We have $\nabla B(\q) = \rho(\o) - \nabla C(\q)$, which represents the instantaneous change of the market maker's loss. For infinitesimal $\epsilon$, let $\q' = \q + \epsilon \left(\rho(\o) - \nabla C(\q)\right)$. Then, 
\[
B(\q') = B(\q) + \nabla B(\q)\cdot \left[\epsilon \left(\rho(\o) - \nabla C(\q)\right)\right]= B(\q) + \epsilon ||\rho(\o) - \nabla C(\q)||^2 \leq B(\q) + \epsilon k^2.
\]
This shows that for any $\q$ we can find a $\q'$ such that the market maker's worst-case loss is at least increased by $\epsilon k^2$. This process can continue for infinite steps. Hence, we conclude that the market maker's loss is unbounded. 
\ignore{
	Consider some outcome $\o$ such that $\rho(\o) \notin \Pi$. Consider the worst-case loss of the market maker for this particular outcome, namely
	\[
		\sup_{\q} \rhob(\o)\cdot \q - C(\q) + C(\0).
	\]
	However, as this optimization is unconstrained over a differentiable function, we see that the condition $\nabla C(\q) = \rho(\o)$ can never be attained, even in the limit, since the set $\{ \nabla C(\q) : \forall \q \} = \Pi$ and $\Pi$ is compact. Hence the optimization is unbounded.
	}
\qed

\subsection{Proof of Proposition~\ref{prop:newworstcaseloss}}

This proof is nearly identical to the proof of Theorem~\ref{thm:worstcaseloss}.  The only major difference is that now $C(\0) = - \min_{\x \in \Pi} R(\x)$ instead of $C(\0) = - \min_{\x \in \H(\rhob(\O))} R(\x)$, but this is equivalent since we have assumed that $\nabla C(\0) \in \H(\rhob(\O))$.  $R(\rhob(\o))$ is still well-defined and finite since we have assumed that $\H(\rhob(\O)) \subseteq \Pi$.

\ignore{
	Let the final quantity vector be denoted $\q_T$ and the final price vector $\x_T$. Let the outcome be $\o$. By the definition of $C$ as the conjugate of $R$, it follows that $C(\0) = - \min_{\x \in \Pi} R(\x)$ and $C(\q_T) = \x_T \cdot \q_T - R(\x_T)$ for some $\x_T \in \Pi$. Hence we can write the loss of the market maker as
	\begin{eqnarray*}
		\q_T \cdot \rhob(\o) - C(\q_T) + C(\0) & = & \q_T \cdot (\rhob(\o) - \x_T) + R(\x_T) - \min_{\x \in \Pi} R(\x) \\
		& \leq & \nabla R(\x_T) \cdot (\rhob(\o) - \x_T) + R(\x_T)  - \min_{\x \in \Pi} R(\x) \\
		& = & - (R(\rhob(\o)) - R(\x_T)  - \nabla R(\x_T) \cdot (\rhob(\o) - \x_T )) + R(\rhob(\o))  - \min_{\x \in \Pi} R(\x) \\ 
		& = & R(\rhob(\o))  - \min_{\x \in \Pi} R(\x) - D_R(\rhob(\o)),\x_T)
	\end{eqnarray*}
	We note that $R(\rhob(\o))$ is well-defined and finite because $\H(\rhob(\O)) \subseteq \Pi$.
	Also, the inequality in the second line follows from the first-order optimality condition for convex optimization:
	\[
		\x = \arg\min_{\x' \in \Pi} f(\x') \quad \implies \quad \nabla f(\x) \cdot (\y - \x) \geq 0 \;\; \text{ for any } \y \in \Pi.
	\]	
Importantly, the inequality is tight when $\nabla C(\q_T) = \x_T$ which is necessarily achievable for some $\q_T$.
	By assumption, the initial price vector $\arg\min_{\x \in \Pi} R(\x)$ is a member of $\H(\rhob(\O))$ which implies that $R(\rhob(\o))  - \min_{\x \in \Pi} R(\x)$ is independent of the choice of $\Pi$. The remaining negative term, $- D_R(\rhob(\o)),\x_T)$, vanishes as $\x_T \to \rhob(\0)$. 
}
\qed

\subsection{Proof of Proposition~\ref{prop:arbitragebound}}

	A trader looking to earn a guaranteed profit when the current quantity is $\q_0$ hopes to purchase a bundle $\r$ so that the worst-case profit $\min_{\o \in \O} \rhob(\o) \cdot \r - C(\q_0 + \r) + C(\q_0)$ is as large as possible. Notice that this quantity is strictly positive since $\r = \0$, which always has 0 profit, is one option. Thus, a trader would like to solve the following objective:
	\begin{eqnarray*}
		& & \max_{\r \in \reals^K} \min_{\o \in \O} \rhob(\o) \cdot \r - C(\q_0 + \r) + C(\q_0) \\
		& = & \min_{\x \in \H(\rhob(\O))} \max_{\r \in \reals^K} \x \cdot \r - C(\q_0 + \r) + C(\q_0) \\
		& = & \min_{\x \in \H(\rhob(\O))} \max_{\r \in \reals^K} \x \cdot(\q + \r) - C(\q_0 + \r) + C(\q_0) - \x \cdot \q_0 \\
		& = & \min_{\x \in \H(\rhob(\O))} R(\x) + C(\q_0) - \x \cdot \q_0 \\
		& = & \min_{\x \in \H(\rhob(\O))} R(\x) + \x_0 \cdot \q_0 - R(\x_0) - \x \cdot \q_0 \\
		& \geq & \min_{\x \in \H(\rhob(\O))} D_R(\x,\x_0).
	\end{eqnarray*}
	The first equality with the $\min/\max$ swap holds via Sion's Minimax Theorem~\cite{sion1958general}. The last inequality was obtained using the first-order optimality condition of the solution $\x_0 = \arg\max_{\x \in \Pi} \x \cdot \q_0 - R(\x)$ for the vector $\x - \x_0$ which holds since $\x \in \Pi$. 
\qed

\end{document}